\documentclass[10pt,a4paper,final]{article}

\usepackage{enumerate}

\usepackage{amssymb,amsmath,amsthm,enumitem}
\usepackage{flushend}
\usepackage{cuted}
\usepackage{lipsum, color}
\usepackage[utf8]{inputenc}
\usepackage[english]{babel}
\usepackage{blindtext}
\usepackage{wrapfig}

\usepackage{graphicx}
\usepackage[breaklinks=true,colorlinks=true,linkcolor=blue,urlcolor=blue,citecolor=blue]{hyperref}
\usepackage{authblk}
\usepackage{sidecap}
\usepackage[cm]{fullpage}

\usepackage{cite}  

\usepackage{xcolor}
\usepackage{soul}
\usepackage{empheq}
\usepackage[many]{tcolorbox}

\tcbset{
	highlight math style={
		colback=yellow,
		arc=0pt,
		outer arc=0pt,
		boxrule=0pt,
		top=2pt,
		bottom=2pt,
		left=2pt,
		right=2pt,
	}
}

\usepackage{multicol} 
\usepackage{caption}
\usepackage{capt-of}
	\sidecaptionvpos{figure}{c}

	\newcommand{\removed}[1]{}
	\newcommand{\replace}[1]{}
	\newcommand{\qprime}{\ensuremath{q^\prime{}}}

	\newcommand{\Eqn}[1]{Eq.~(\ref{#1})}

	\newcommand{\Fig}[1]{Figure~\ref{#1}}

	\newcommand{\quotes}[1]{``#1''}
	
\def\Plus{\texttt{+}}
\def\Minus{\texttt{-}}

\begin{document}
\onecolumn

\title{Generalised ballooning theory of two dimensional tokamak modes}

\author[1,2,3]{P. A. Abdoul}
\author[3]{D. Dickinson}
\author[4]{C. M. Roach}
\renewcommand\Authands{ and }
\author[3]{H. R. Wilson}
\affil[1]{Charmo University, College of Education, Department of Physics, 46023 Chamchamal/Al-Sulaimaniyah, Kurdistan region, F.R. Iraq}
\affil[2]{University of Sulaimani, College of Science, Department of Physics, 46001 Al-Sulaimaniyah, Kurdistan region, F.R. Iraq}
\affil[3]{York Plasma Institute, Department of Physics, University of York, Heslington, York, YO10 5DD, UK}
\affil[4]{CCFE, Culham Science Centre, Abingdon, Oxforshire, OX14 3DB, UK}


\maketitle

\begin{abstract}
\label{abstract}

In this work, using solutions from a local gyrokinetic flux-tube code combined with higher order ballooning theory, a new analytical approach is developed to reconstruct the global linear mode structure with associated global mode frequency. In addition to the isolated mode (IM), which usually peaks on the outboard mid-plane, the higher order ballooning theory has also captured other types of less unstable global modes: \textbf{(a)} the weakly asymmetric ballooning theory (WABT) predicts a mixed mode (MM) that undergoes a small poloidal shift away from the outboard mid-plane, \textbf{(b)} a relatively more stable general mode (GM) balloons on the top (or bottom) of the tokamak plasma. In this paper, an analytic approach is developed to combine these disconnected analytical limits into a single generalised ballooning theory (GBT). This is used to investigate how an IM behaves under the effect of sheared toroidal flow. For small values of flow an IM initially converts into a MM where the results of WABT are recaptured, and eventually, as the flow increases, the mode asymptotically becomes a GM on the top (or bottom) of the plasma. This may be an ingredient in models for understanding why in some experimental scenarios, instead of large edge localised modes(ELMs), small ELMs are observed. Finally, our theory can have other important consequences, especially for calculations involving Reynolds stress driven intrinsic rotation through the radial asymmetry in the global mode structures. Understanding the intrinsic rotation is significant because external torque in a plasma the size of ITER is expected to be relatively low.

\end{abstract}


\section{Introduction}
\label{Intro_Theory}
Cross-field turbulent transport of both heat and particles, widely believed to be caused by low frequency microinstabilities is a main obstacle to achieving ignition in magnetically confined plasmas \cite{Horton1999}. Hence, it is crucial to understand them and find a way to reduce their effects. Among the most interesting subject in this area, is the impact of global effects, such as plasma profile variations \cite{Peshwaz2015, Camenem2011, Hill2012}. In this work, we focus on axisymmetric tokamaks in which the spatial component of the linearized gyrokinetic equations describing microinstabilities can be reduced to a 2D eigenmode equation in radius, $x$, and poloidal angle, $\theta$. There are two main numerical treatments of the linear gyrokinetic equations, namely global and local, as we now discuss. 

A global solution takes into account the effect of the radial profile variations, involving full solutions to the 2D eigenmode equations, providing both the global mode structure and associated global mode frequency. By contrast, the local approach employs a so-called ballooning theory to reduce the problem to a 1-D local system \cite{Connor1978, Connor1979, Hastie1979, Pegoraro1981}. Moreover, ballooning theory has proven to be a powerful tool to investigate high toroidal mode number, $n \gg 1$, instabilities in tokamak plasmas. For this type of mode the distance between two neighboring rational surfaces is negligible compared to the equilibrium scale length. Therefore, one can take advantage of this scale separation and employ the WKB treatment in a perturbation expansion. The lowest order in $1/n$, equivalent to local ballooning theory, exploits a so-called translation (or ballooning) symmetry, i.e the rational surfaces associated with an instability approximately experience the same equilibrium parameters as $n \rightarrow \infty$. This theory provides both local mode structures along field lines and the associated local complex mode frequency \cite{Connor1978}. However, proceeding to the next order in the expansion breaks the translational symmetry, providing constraints to the local solutions, which determine the 2D global mode structures and their global mode frequency \cite{Connor1978, Connor1993, Taylor1996, Xie2012, David2014}. In this paper, in the context of the higher order \quotes{ballooning formalism}, a new analytical theory is presented to reconstruct the 2D mode structures and calculate their mode frequencies from solutions to the local 1D ballooning equations. There are two main advantages of using local results to build up global mode structures; \textbf{(a)} they are computationally less intensive compared to full global solutions and \textbf{(b)} we can gain a deeper understanding of the physics behind the global mode structures \cite{Peshwaz2015, PeshwazPhD, DavidPhD}. 

The global mode structure can be reconstructed purely from the solutions to the local ballooning equations using the Fourier-Ballooning representation \cite{Mahajan1991}:

	\begin{equation}
	\begin{aligned}
\label{equ_FB}
\tilde{\phi}(x,\theta) = \int_{\Minus\infty}^{\Plus\infty}\xi(x, \theta_{0}, \theta) \exp\bigg[-inq_{0}\theta\bigg] 
\exp\bigg[-in\qprime x\left(\theta-\theta_{0}\right)\bigg]Y(\theta_{0}) d\theta_{0}
\end{aligned}
\end{equation}
Here, $\tilde{\phi}(x,\theta)$ is the global mode. The function $\xi(x, \theta_{0},\theta)$ represents the local mode structure, and it is invariant under the double transformation $\theta_{0}\rightarrow \theta_{0} + 2\pi N$ and $\theta \rightarrow \theta+2\pi N$ for integer $N$. The poloidal angle $\theta$ measures the distance along the magnetic field, $n$ is the toroidal mode number and $q_{0}$ and $\qprime=dq/dx$ are the value of safety factor and its derivative on a rational surface at $x=0$, respectively. The radial distance from a reference rational surface, $x=(r-r_{s})/a$, is  normalised to the tokamak minor radius $a$. It is assumed that the envelope $Y(\theta_{0})$ varies much more rapidly than the local mode structure $\xi(\theta_{0})$ with $\theta_{0}$ (i.e $\frac{dY(\theta_{0})}{d\theta_{0}} \gg \frac{d\xi(\theta_{0})}{d\theta_{0}}$) and satisfies the following differential equation \cite{Peshwaz2015}:
\begin{eqnarray}
	\begin{aligned}
\label{equ_YODE}
\frac{\lambda_{xx}(0,\theta_{0})}{2n^2\qprime^2}\frac{d^{2} Y(\theta_{0})}{d\theta_{0}^{2}} - \frac{i\lambda_{x}(0,\theta_{0})}{n\qprime}\frac{d Y(\theta_{0})}{d\theta_{0}} +  
\bigg[\Omega - \lambda (0,\theta_{0}) \bigg] Y(\theta_{0}) = 0
\end{aligned}
\end{eqnarray}
 Where, the periodic functions $\lambda(0,\theta_{0})$, $\lambda_{x}(0,\theta_{0})$ and $\lambda_{xx}(0,\theta_{0})$ are obtained from the Taylor series expansion of the local complex mode frequency $\lambda(x,\theta_{0}) = \lambda_{r}(x,\theta_{0}) + i \lambda_{i}(x,\theta_{0})$ about $x=0$ ($\lambda_{x}$ and $\lambda_{xx}$ denote the first and second derivative with respect to $x$). Here, the real $\lambda_{r}(x,\theta_{0})$ and imaginary $\lambda_{i}(0,\theta_{0})$ components correspond to the local frequency and local growth rate, respectively and  $\Omega = \omega + i \gamma$ is the global complex mode frequency with $\omega$ and $\gamma$ being the global real frequency and growth rate. Note that, the periodicity constraint on $\tilde{\phi}(x,\theta)$ with $\theta$ requires a periodic $Y(\theta_{0})$ in $\theta_{0}$. Having knowledge of $Y(\theta_{0})$ and its eigenvalue $\Omega$ from \Eqn{equ_YODE}, together with the local mode structure $\xi(x, \theta_{0}, \theta)$, obtained from the local gyrokinetic codes, one may reconstruct the 2D mode structure, $\tilde{\phi}(x,\theta)$, from \Eqn{equ_FB}.

Providing a full analytic solution for \Eqn{equ_YODE} is quite challenging, but it has been solved in a few limits \cite{Connor1978,  Connor1993, Taylor1996, David2014, Xie2012, Xie2016}; assuming $\lambda_{x}(0,\theta_{0}) = 0$, corresponding to a special case where $\lambda(x,\theta_{0})$ has a stationary point at $x=0$, leads to a highly unstable isolated mode (IM) that, for the poloidally up-down symmetric plasma equilibrium typically sits on the outboard mid-plane with global growth rate $\gamma$ obtained from $\Omega = \lambda(x=0,\theta_{0}=0) + O(1/n)$. However, when $\lambda_{x}(0,\theta_{0}) \ne 0$ and the second term in \Eqn{equ_YODE} dominates the first term involving $\lambda_{xx}(0,\theta_{0})$, then the latter term can be neglected and one finds a relatively less unstable general mode (GM) that balloons on the top or bottom of the tokamak plasma. Its global growth rate and frequency involves an average over a period of $\theta_{0}$, for instance see \Eqn{equ_plus_GMomg0} (specifically $\Omega = \oint{\left[\lambda(0,\theta_{0})\right] d\theta_{0}}$). On the other hand, the weakly asymmetric ballooning theory (or WABT) expands \Eqn{equ_YODE} about $\theta_{0}=0$ and, by retaining up to the second order term in the expansion, leads to a mixed mode (MM), that slightly shifts with respect to the outboard mid-plane \cite{PeshwazPhD, Xie2016}\footnote{Note that WABT in \cite{PeshwazPhD} takes into account the variation of $\lambda_{x}(0,\theta_{0})$ and $\lambda_{xx}(0,\theta_{0})$ with $\theta_{0}$. While \cite{Xie2016} is limited to a special case where the weakly $\theta_{0}$-dependent part of $\lambda_{x}(0,\theta_{0})$ and $\lambda_{xx}(0,\theta_{0})$ are neglected. This simplification reduces \Eqn{equ_YODE} to the well known Weber equation which is exactly solved in terms of Hermite polynomials.}. More usually in realistic experimental situations, the global modes can sit anywhere in the poloidal plane \cite{Peshwaz2015, Camenem2011}. Therefore, to account for these modes, we have presented a new analytical theory that combines all the aforementioned analytical solutions into a single theory, namely a generalised ballooning theory (GBT). Specifically we can capture the sudden transition of an IM, through MM, into a GM as equilibrium profiles, such as the flow shear, evolve. This inter-mode transition has been considered as the basis for a model that could be used to understand the mechanism that underlies some classes of small ELMs \cite{David2014, Arka2016} and, hence, our theory can provide additional physical insights into these models. 

Furthermore, it is worth mentioning that $\lambda_{x}(0,\theta_{0})$ incorporates the effect of profile shearing, a special case of which is the rotational flow shear corresponding to $[\lambda_{x}(0,\theta_{0})]_{i}=0$, but $[\lambda_{x}(0,\theta_{0})]_{r} \ne 0$, where the subscripts $r$ and $i$ correspond to the real and imaginary components, respectively. Flow shear can suppress the microinstabilities, thereby improving the plasma confinement \cite{Connor2007, Terry2000, Kishimoto1999, Roach2009, Biglari1999, Waltz1994}. However, it has recently been demonstrated that, in the presence of profile variations, flow shear can also destabilise the global instability such that for a critical value of flow shear, where the flow precisely compensates the effect of the profile variations, a highly unstable IM is captured \cite{Hill2012, Peshwaz2015}. Finally, our theory has revealed that any poloidal shifts away from the outboard mid-plane are always accompanied by an asymmetry in the radial eigenmode structure\footnote{Note that, by taking the radial slice from the constructed mode structure at $\theta=0$, the radial asymmetry is measured with respect to a rational surface about which the mode peaks.}. This radial asymmetry can generate a so-called Reynolds stress \cite{Zhang1992, Xie2012, XieZhang2016}, which is very important, especially in calculations that employ quasilinear theory to model intrinsic rotation arising from Reynolds stress. Intrinsic rotation may be beneficial for a machine like ITER for which external momentum sources are weak or not practical. In such theories Reynolds stress can, in principle, contribute to the generation of poloidal flows in the edge region of tokamak plasmas during the low to high (L-H) mode transition \cite{Zhang1995, Burrell1997}.

This paper is organised as follows. Section \ref{solutions_analytic} is devoted to solving \Eqn{equ_FB} and \Eqn{equ_YODE} analytically. In this section, we have applied our theory to explain and understand the radial asymmetry associated with a mode that undergoes a poloidal shift with respect to the outboard mid-plane. In Section \ref{Validation} we have validated our theory by recapturing an IM, MM and GM in special limits. In Section \ref{GBT_Application}, using available data from the literature for a toroidal ion temperature gradient (ITG) model in a large aspect ratio circular tokamaks, we have benchmarked our calculations against numerical solutions. Finally, the conclusion and future plans are presented in Section \ref{conclusions}.

\section{Generalised ballooning theory}
\label{solutions_analytic}
In this section, we develop a new analytical approach that combines all previously obtained analytical solutions for IMs, MMs and GMs into a single generalised ballooning theory (GBT).

\subsection{The envelope \texorpdfstring{$Y(\theta_{0})$}{Lg} and \texorpdfstring{$\Omega$}{Lg}}
\label{Aalytic_Y_Om}
We start with \Eqn{equ_YODE} and seek localised solutions with a form:
\begin{eqnarray}
\label{equ_YthF}
Y(\theta_{0}) = \exp{\left[-n\qprime F(\theta_{0})\right]}
\end{eqnarray}
where $F(\theta_{0})$ is a $2\pi$ periodic function of $\theta_{0}$. In this paper, assuming $n\qprime \gg 1$, we shall generalise the restricted solutions of weak asymmetric ballooning theory (WABT) by expanding about $\theta_{0}=\theta_{b}$ rather than $\theta_{0}=0$. At this point $\theta_{b}$ is arbitrary but in the following we shall describe the procedure to determine its value. We now Taylor expand about $\theta_{0}=\theta_{b}$, retaining up to the second order terms in $\theta_{0}-\theta_{b}$, and write\footnote{It is worth mentioning that we are evaluating $Y(\theta_{0})$ only for real $\theta_{0}$ but the parameter $\theta_{b}$ can be, in general, complex. The imaginary component of $\theta_{b}$ is related to the symmetry of $Y(\theta_{0})$ which, in turn, as we shall see in \ref{Analytic_Eigenmode_Structure}, causes a radial shift of the mode off $x=0$.}
\begin{eqnarray}
\label{equ_Yth0}
Y(\theta_{0}) \approx \exp{\left[-n\qprime\alpha \left(\theta_{0}-\theta_{b}\right)^{2}\right]}
\end{eqnarray}
\begin{eqnarray}
\begin{aligned}
\label{equ_taylor0}
\lambda_{xx}(0,\theta_{0}) &\approx \lambda_{xx}(0,\theta_{b}) + \lambda_{xx\theta_{0}}(0,\theta_{b})\left(\theta_{0} -\theta_{b}\right) 
+ \frac{\lambda_{xx\theta_{0}\theta_{0}}(0,\theta_{b})}{2}\left(\theta_{0}-\theta_{b}\right)^{2}
\\
\lambda_{x}(0,\theta_{0}) &\approx \lambda_{x}(0,\theta_{b}) + \lambda_{x\theta_{0}}(0,\theta_{b})\left(\theta_{0} -\theta_{b}\right) 
+ \frac{\lambda_{x\theta_{0}\theta_{0}}(0,\theta_{b})}{2}\left(\theta_{0}-\theta_{b}\right)^{2}
\\
\lambda(0,\theta_{0}) &\approx \lambda(0,\theta_{b}) + \lambda_{\theta_{0}}(0,\theta_{b})\left(\theta_{0} -\theta_{b}\right) 
+ \frac{\lambda_{\theta_{0}\theta_{0}}(0,\theta_{b})}{2}\left(\theta_{0}-\theta_{b}\right)^{2}
\end{aligned}
\end{eqnarray}
where, the subscripts $\theta_{0}$ and $\theta_{0}\theta_{0}$ on $F$ and the coefficients $\lambda$, $\lambda_{x}$ and $\lambda_{xx}$ refer, respectively, to their first and second derivatives with respect to $\theta_{0}$ and are evaluated at $\theta_{0} = \theta_{b}$. Here we have absorbed constant factors into $Y(\theta_{0})$ and $F_{\theta_{0}\theta_{0}}(\theta_{b})$ is replaced by $2\alpha$. In addition, we define $\theta_{b}$ as the location of maximum $F(\theta_{0})$, so that $F_{\theta_{0}}(\theta_{b})=0$. After substituting \Eqn{equ_Yth0} and \Eqn{equ_taylor0} back into \Eqn{equ_YODE}, and equating coefficients with like powers of $(\theta_{0} - \theta_{b})$, we obtain the following equation for $\alpha$:
\begin{eqnarray}
\label{equ_alpha}
\begin{aligned}
\alpha = \frac{\lambda_{xx\theta_{0}\theta_{0}}(0,\theta_{b})}{8n\qprime\lambda_{xx}(0,\theta_{b})}
				- \frac{i\lambda_{x\theta_{0}}(0,\theta_{b})}{2\lambda_{xx}(0,\theta_{b})}
+ \delta \sqrt{\left(\frac{\lambda_{xx\theta_{0}\theta_{0}}(0,\theta_{b})}{8n\qprime\lambda_{xx}(0,\theta_{b})} - \frac{i\lambda_{x\theta_{0}}(0,\theta_{b})}{2\lambda_{xx}(0,\theta_{b})}\right)^{2} 
+ \frac{\lambda_{\theta_{0}\theta_{0}}(0,\theta_{b})}{4\lambda_{xx}}}, 
\end{aligned}
\end{eqnarray}
an equation to be solved for $\theta_{b}$
\begin{equation}
\begin{aligned}
\label{equ_thetab}
\bigg[2i\lambda_{x}(0,\theta_{b}) - \frac{\lambda_{xx\theta_{0}}(0,\theta_{b})}{n\qprime}\bigg] \alpha
		= \lambda_{\theta_{0}}(0,\theta_{b}),
\end{aligned}
\end{equation}
and the following solution for $\Omega$:
\begin{eqnarray}
\label{equ_omg}
\begin{aligned}
	\Omega	= \lambda(0,\theta_{b}) 
		+ \bigg[\frac{\lambda_{xx}(0,\theta_{b})}{n\qprime}\bigg] \alpha
\end{aligned}
\end{eqnarray}
where $\delta =\pm 1$ and the sign is chosen such that $Y(\theta_{0})$ is localised in $\theta_{0}$ space. Finally, these results can be combined to derive the following form for $Y(\theta_{0})$
\begin{eqnarray}
\label{equ_Yth0abs}
\begin{aligned}
Y(\theta_{0}) = \exp{\bigg[-n\qprime\alpha(\theta_{0}-\theta_{m})^2\bigg]}   
\exp{\bigg[\left(\frac{2in\qprime|\alpha|^{2}}{\alpha_{r}}\theta_{bi}\right) \theta_{0}\bigg]},
\end{aligned}
\end{eqnarray}
where, constant factors are absorbed into $Y(\theta_{0})$ and the subscripts $r$ and $i$ corresponding to the real and imaginary parts, respectively. The real parameter,
\begin{equation}
\label{equ_thetam}
\begin{aligned}
\theta_{m} = \theta_{br} - \frac{\alpha_{i}}{\alpha_{r}}\theta_{bi} 
\end{aligned}
\end{equation}
determines where the magnitude of $Y(\theta_{0}$) peaks in $\theta_{0}$ space\footnote{Note that, $\theta_{m}$ also determines the mode physical poloidal position with respect to the outboard mid-plane.}. Furthermore, the last exponential term on the right hand side of \Eqn{equ_Yth0abs} controls the symmetry of $|Y(\theta_{0})|$ about a line that goes through $\theta_{0} = \theta_{m}$. We note that the resultant $Y(\theta_{0})$ is symmetric about $\theta_{0} = \theta_{m}$ only for a set of equilibrium parameters that provides $\theta_{bi} = 0$, corresponding to $\theta_{m}=\theta_{br}$ in \Eqn{equ_thetam}. This symmetry breaking has important consequences for the reconstructed global mode structure, as we now turn to discuss in the following subsection.

\subsection{The global mode structure \texorpdfstring{$\tilde{\phi}(x,\theta)$}{Lg}}
\label{Analytic_Eigenmode_Structure}
Now to calculate the global eigenmode structures, we substitute $Y(\theta_{0})$ from \Eqn{equ_Yth0} into the Fourier ballooning representation in \Eqn{equ_FB}, and assuming that $\xi(x,\theta,\theta_{0})$ varies slowly with $\theta_{0}$ compared to $Y(\theta_{0})$, we can then apply the stationary phase approximation to the integral in \Eqn{equ_FB}. To lowest order we find:
	\begin{eqnarray}\label{equ_GaussInt}
	\begin{aligned}
\tilde{\phi}(x,\theta)  \approx \xi(x,\theta_{m},\theta)   
\exp\bigg[-n\qprime\alpha\theta_{b}^{2}\bigg] \exp\bigg[-inq_{0}\theta\bigg]
\exp\left[-in\qprime\theta x - \frac{n\qprime}{4\alpha}\left(x-2i\alpha\theta_{b}\right)^{2}\right] 
\\ 
\times
\int_{-\infty}^{+\infty}  \exp \left[-n\qprime\alpha\left(\theta_{0}-\left(\theta_{b}+\frac{ix}{2\alpha}\right)\right)^{2} \right]
 d\theta_{0}
\end{aligned}
\end{eqnarray}
where $\theta_{m}$ is defined in \Eqn{equ_thetam}. The integration on the right hand side of \Eqn{equ_GaussInt} is a well known Gaussian with a complex shift $\theta_{b}+\frac{ix}{2\alpha}$ and the result is $\sqrt{\frac{\pi}{n\qprime\alpha}}$ \cite{Kneser1958, Remmert1991}. To determine the mode physical radial shift and investigate its radial symmetry we shall first decompose $\alpha$ and $\theta_{b}$ into their real and imaginary components to obtain
\begin{eqnarray}
\label{equ_phix1}
\begin{aligned}
\tilde{\phi}(x,\theta) \approx \sqrt{\frac{\pi}{n\qprime\alpha}} \xi(x, \theta_{m}, \theta) \exp\bigg[-inq_{0}\theta\bigg] 
 \exp\bigg[-in\qprime\theta x\bigg]
\exp\bigg[-\frac{n\qprime\alpha_{r}}{4|\alpha|^2}\left(x^2 - 2x_{m}x\right)\bigg] 
\\ \times
 \exp\bigg[\frac{in\qprime\alpha_{i}}{4|\alpha|^2}\left(x^2 - 2x_{m}x\right)\bigg] 
\exp\bigg[in\qprime\theta_{m} x\bigg]
\end{aligned}
\end{eqnarray}
Now by completing the square in $x$ then rearrange the terms in \Eqn{equ_phix1} we get 
	\begin{eqnarray}
\label{equ_phixm}
\begin{aligned}
\tilde{\phi}(x,\theta) 
\approx  \xi(x, \theta_{m}, \theta) \exp\bigg[-inq_{0}\theta\bigg] 
 \exp\bigg[-\frac{n\qprime}{4\alpha}\left(x-x_{m}\right)^{2}\bigg] 
\exp\bigg[-in\qprime\left(\theta - \theta_{m}\right)x\bigg]
\end{aligned}
\end{eqnarray}
where, again we have absorbed constant factors into $\tilde{\phi}(x,\theta)$. Here, the real parameter,
	\begin{eqnarray}
\label{equ_xm}
\begin{aligned}
x_{m}  =  - \frac{2 \left|\alpha\right|^{2}}{\alpha_{r}} \theta_{bi}
\end{aligned}
\end{eqnarray}
represents a physical radial shift away from $x=0$. From \Eqn{equ_phixm} and \Eqn{equ_xm} it is clear that if $\theta_{bi} \ne 0$, corresponding to asymmetric $Y(\theta_{0})$ about $\theta_{0} = \theta_{m}$ in \Eqn{equ_Yth0abs}, the reconstructed global mode, in turn, undergoes a radial shift away from its associated rational surface at $x=0$. Furthermore, from the last exponential on the right side of \Eqn{equ_phixm}, we can see that $\tilde{\phi}(x,\theta)$ has a radial symmetry about $x=x_{m}$ only when $\theta = \theta_{m}$. However, if we take the radial slice from the constructed mode structure at $\theta=0$, only IMs, for which $\theta = \theta_{m}=0$, have radial symmetry. Thus, we conclude that; any poloidal shift with respect to $\theta=0$ introduces asymmetry into the radial mode structures.

Finally, the mode's radial width is determined from the full width at half maximum (FWHM) of the Gaussian (second exponential term on the right hand side in \Eqn{equ_phixm}), which reads
	\begin{eqnarray}
\label{equ_dx}
\Delta_{x} = \frac{4 \sqrt{log(2)} |\alpha|}{\sqrt{n|\qprime|}\sqrt{|\alpha_{r}|}}
\end{eqnarray}

\section{Validation of GBT}
\label{Validation}
The core purpose of this section is to test some of the key predictions of our new theory by reproducing analytical results for all IMs, MMs and GMs that have been previously derived in \cite{Connor1993, Taylor1996, Xie2012, Xie2016} for instance. We note that, to obtain theses modes, a few simplifications and assumptions are made in literature. Hence, to recapture those solutions from our GBT theory, we shall employ similar simplifications. Firstly, we neglect the weak $\theta_{0}$-dependence of both $\lambda_{x}(0,\theta_{0})$ and $\lambda_{xx}(0,\theta_{0})$ and write
\begin{equation}
\label{equ_lambdaxmodel0}
\begin{aligned}
\lambda_{x}(0,\theta_{0})  & = \lambda_{x}
\\
\lambda_{xx}(0,\theta_{0}) & = \lambda_{xx}
\end{aligned}
\end{equation}
where $\lambda_{x}$ and $\lambda_{xx}$ are, in general, complex numbers. Moreover, knowing the fact that $\lambda(0,\theta_{0})$ is periodic in $\theta_{0}$ we may Fourier expand to obtain
\begin{equation}
\label{equ_lambdamodel}
\lambda(0,\theta_{0}) = \lambda_{0} + \lambda_{1}\cos{(\theta_{0})} 
\end{equation}
where, only two Fourier harmonics have been retained. For this simplified model \Eqn{equ_alpha} reduces to
\begin{eqnarray}
\label{equ_alphamodel}
\begin{aligned}
\alpha = \frac{i\delta}{2}\sqrt{\frac{\lambda_{1}}{\lambda_{xx}}\cos{(\theta_{b})}} 
\end{aligned}
\end{eqnarray}
This, in turn, simplifies both $Y(\theta_{0})$ in \Eqn{equ_Yth0} and $\Omega$ in \Eqn{equ_omg}:
\begin{eqnarray}
\label{equ_Ymodel}
\begin{aligned}
Y(\theta_{0}) = \exp{\left[\frac{-in\qprime\delta}{2}\sqrt{\frac{\lambda_{1}}{\lambda_{xx}}\cos{(\theta_{b})}} \bigg(\theta_{0} - \theta_{b}\bigg)^{2}\right]}
\end{aligned}
\end{eqnarray}
and
\begin{eqnarray}
\label{equ_omgmodel}
\begin{aligned}
	\Omega	= \lambda_{0} + \lambda_{1} \cos{(\theta_{b})} + \frac{i\lambda_{xx}\delta}{2n\qprime} \sqrt{\frac{\lambda_{1}}{\lambda_{xx}}\cos{(\theta_{b})}}  
\end{aligned}
\end{eqnarray}
Furthermore, the corresponding global mode structure $\tilde{\phi}(x,\theta)$ (from \Eqn{equ_phixm}) with its radial mode width (\Eqn{equ_dx}) are, respectively, reduced to 
\begin{equation}\label{equ_phixmmodel}
\begin{aligned}
\tilde{\phi}(x,\theta)  \sim \xi(x, \theta_{m}, \theta) \exp\bigg[-inq_{0}\theta\bigg] 
 \exp\bigg[\frac{in\qprime\delta}{2\sqrt{\frac{\lambda_{1}}{\lambda_{xx}}\cos{(\theta_{b})}} }\bigg(x-x_{m}\bigg)^{2}\bigg]
 \exp\bigg[-in\qprime\left(\theta - \theta_{m}\right)x\bigg]
\end{aligned}
\end{equation}
and
\begin{equation}\label{equ_dxmodel}
\begin{aligned}
\Delta_{x} &= \frac{2 \sqrt{2log(2)}}{\sqrt{n|\qprime|}}  
\frac{\left|\sqrt{\frac{\lambda_{1}}{\lambda_{xx}}\cos{(\theta_{b})}}\right|}{\sqrt{\left|\left[\sqrt{\frac{\lambda_{1}}{\lambda_{xx}}\cos{(\theta_{b})}}\right]_{i}\right|}}
\end{aligned}
\end{equation}
where, both $\theta_{m}$ (from \Eqn{equ_thetam}) and $x_{m}$ (from \Eqn{equ_xm}), respectively, take the following forms
\begin{eqnarray}
\label{equ_thetammodel}
\begin{aligned}
\theta_{m} = \theta_{br} + \frac{ \bigg[\sqrt{\frac{\lambda_{1}}{\lambda_{xx}} \cos{(\theta_{b})}} \bigg]_{r} } { \bigg[\sqrt{\frac{\lambda_{1}}{\lambda_{xx}} \cos{(\theta_{b})}} \bigg]_{i} } \times \theta_{bi}
\end{aligned}
\end{eqnarray}
\begin{eqnarray}
\label{equ_xmmodel}
\begin{aligned}
x_{m} =\delta \frac{\left|\sqrt{\frac{\lambda_{1}}{\lambda_{xx}} \cos{(\theta_{b})}} \right|^{2}}{\bigg[\sqrt{\frac{\lambda_{1}}{\lambda_{xx}} \cos{(\theta_{b})}} \bigg]_{i}} \times \theta_{bi}
\end{aligned}
\end{eqnarray}
Now, to calculate $\theta_{b}$, we substitute \Eqn{equ_lambdaxmodel0} and \Eqn{equ_lambdamodel} into \Eqn{equ_thetab} and rearrange to obtain
\begin{equation}
\begin{aligned}
\label{equ_thetabmodel2}
\cos^{2}{(\theta_{b})} + \bigg(\left|\frac{\lambda_{x}^{2}}{\lambda_{1}\lambda_{xx}}\right| \exp{\left[i\arg\left(\frac{\lambda_{x}^{2}}{\lambda_{1}\lambda_{xx}}\right)\right]}\bigg)
\cos{(\theta_{b})} - 1 = 0
\end{aligned}
\end{equation}
where, we have expressed the complex number $\lambda_{x}^{2}/\lambda_{1}\lambda_{xx}$ in polar form to write
\begin{equation}
\label{equ_thetabmodel3}
\frac{\lambda_{x}^{2}}{\lambda_{1}\lambda_{xx}} = \left|\frac{\lambda_{x}^{2}}{\lambda_{1}\lambda_{xx}}\right| \exp{\left[i\arg\left(\frac{\lambda_{x}^{2}}{\lambda_{1}\lambda_{xx}}\right)\right]}
\end{equation}
and \Eqn{equ_thetabmodel2} is written in a form such that the real parameter $\left|\lambda_{x}^{2}/\lambda_{1}\lambda_{xx}\right|$ is used in the subsequent sections to solve \Eqn{equ_thetabmodel2} perturbativley to reproduce the analytical solutions for IMs, MMs and GMs, respectively. Here, $\left|\lambda_{x}^{2}/\lambda_{1}\lambda_{xx}\right|$ and $\arg\left(\lambda_{x}^{2}/\lambda_{1}\lambda_{xx}\right)$ refer to the magnitude and argument of $\lambda_{x}^2/\lambda_{1}\lambda_{xx}$, respectively.

\subsection{Isolated modes: \texorpdfstring{$\lambda_{x} = 0$}{Lg}}
\label{sec_IM}
We start by substituting $\lambda_{x}=0$ into \Eqn{equ_thetabmodel2} to obtain
\begin{equation}
\label{equ_IMthetab}
\cos{(\theta_{b}^{\pm})} = \pm 1
\end{equation}
where plus and minus signs has solutions $\theta_{b}^{+} = 0$ and $\theta_{b}^{-}=\pm \pi$, respectively. Therefore, for both solutions $\theta_{bi}=0$ which then provides a symmetric $Y(\theta_{0})$ in \Eqn{equ_Ymodel} and from \Eqn{equ_thetammodel} we have $\theta_{m}^{+}=\theta_{br}^{+}=0$ (or $\theta_{m}^{-}=\theta_{br}^{-}=\pm\pi$) leading to a mode that peaks on the outboard (or inboard) mid-plane and radially centered on $x=x_{m}^{\pm}=0$. The corresponding analytical solutions, using the results given in section \ref{Validation} (from \Eqn{equ_Ymodel} to \Eqn{equ_dxmodel}), are reduced to the following forms\footnote{Note that $\delta^{+}=\pm 1$ ($\delta^{-}=\pm 1$) and the sign is chosen such that the mode on outboard (inboard) mid-plane is localised about $\theta_{0}=0$ ($\theta_{0}=\pm\pi$).}
\begin{equation}\label{equ_IMY}
\begin{aligned}
Y_{I}^{+}(\theta_{0}) &= \exp{\left[-\frac{in\qprime\delta^{+}}{2}\sqrt{\frac{\lambda_{1}}{\lambda_{xx}}} \left(\theta_{0}\right)^{2}\right]}
\\
Y_{I}^{-}(\theta_{0}) &= \exp{\left[\frac{n\qprime\delta^{-}}{2}\sqrt{\frac{\lambda_{1}}{\lambda_{xx}}} \left(\theta_{0}\mp\pi\right)^{2}\right]}
\end{aligned}
\end{equation}
\begin{equation}\label{equ_IMom}
\begin{aligned}
\Omega_{I}^{\pm} &= \lambda_{0} \pm \lambda_{1} + \frac{i\delta^{\pm}\sqrt{\pm\lambda_{1}\lambda_{xx}}}{2n\qprime}
\end{aligned}
\end{equation}
\begin{equation}\label{equ_IMphi}
\begin{aligned}
\tilde{\phi}_{I}^{+}(x,\theta) \sim \xi(x, \theta_{m}^{+} = 0, \theta) \exp\bigg[-inq\theta\bigg] 
 \exp\bigg[\frac{in\qprime \delta^{+}}{2\sqrt{\frac{\lambda_{1}}{\lambda_{xx}}}}x^{2}\bigg] 
\\
\tilde{\phi}_{I}^{-}(x,\theta) \sim \xi(x, \theta_{m}^{-} = \pm\pi, \theta) \exp\bigg[-inq\theta\bigg] 
 \exp\bigg[\frac{n\qprime \delta^{-}}{2\sqrt{\frac{\lambda_{1}}{\lambda_{xx}}}}x^{2}\bigg] 
\exp\bigg[\pm i\pi n\qprime x\bigg]
\end{aligned}
\end{equation}
\begin{equation}\label{equ_IMdx}
\begin{aligned}
\Delta_{xI}^{\pm} &= \frac{2 \sqrt{2log(2)}}{\sqrt{n|\qprime|}}  
\frac{\left|\sqrt{\pm\lambda_{1}/\lambda_{xx}}\right|}{\sqrt{\left|\left[\sqrt{\pm\lambda_{1}/\lambda_{xx}}\right]_{i}\right|}}
\end{aligned}
\end{equation}
where $q=q_{0} + \qprime x$ and the subscript \quotes{I} refers to an isolated mode. Furthermore, our \Eqn{equ_IMY} and \Eqn{equ_IMom} corresponding to $Y(\theta_{0})$ and its eigenvalue $\Omega$, respectively, are exactly equivalent to Eqs. (4) and (5) in \cite{David2014} (note that in \cite{David2014} our coefficients $\lambda_{1}$ and $\lambda_{xx}$ are replaced by $-\Omega_{\theta_{0}\theta_{0}}$ and $\Omega_{xx}$, respectively). Hence, we conclude that the mode that sits in the bad curvature region at $\theta=0$, corresponding to plus sign solutions, is an IM. However, the additional mode that has been captured on the good curvature region at $\theta =\pm \pi$, corresponding to minus sign solutions, we called an anti-IM. Moreover, for a real physical system we expect IMs to be more unstable than anti-IMs due to the stabilising influence of good curvature region. This, in turn, provides a constraint on the model coefficients. To see this, we may use \Eqn{equ_IMom} and define a controlling parameter $\Delta_{\Omega}^{\pm}$ such that $\Delta_{\Omega}^{\pm} = \Omega_{I}^{+} - \Omega_{I}^{-}$; after neglecting small corrections arising from $O(1/n)$ terms we obtain $\Delta_{\Omega}^{\pm} \approx  2\lambda_{1}$. Note, therefore, that an IM can be more unstable than an anti-IM if and only if
\begin{equation}\label{equ_IMDelom}
\begin{aligned}
\left[\Delta_{\Omega}^{\pm}\right]_{i} = 2\left[\lambda_{1}\right]_{i}> 0
\end{aligned}
\end{equation}
Here, subscript $i$ refer to the imaginary component.

\subsection{Mixed Modes: \texorpdfstring{$\left|\lambda_{x}^{2}/\lambda_{1}\lambda_{xx}\right| \ll 1$}{Lg}}
\label{sec_MM}
\begin{figure*}[!t]
	\centering
	\includegraphics[width=6.0in]{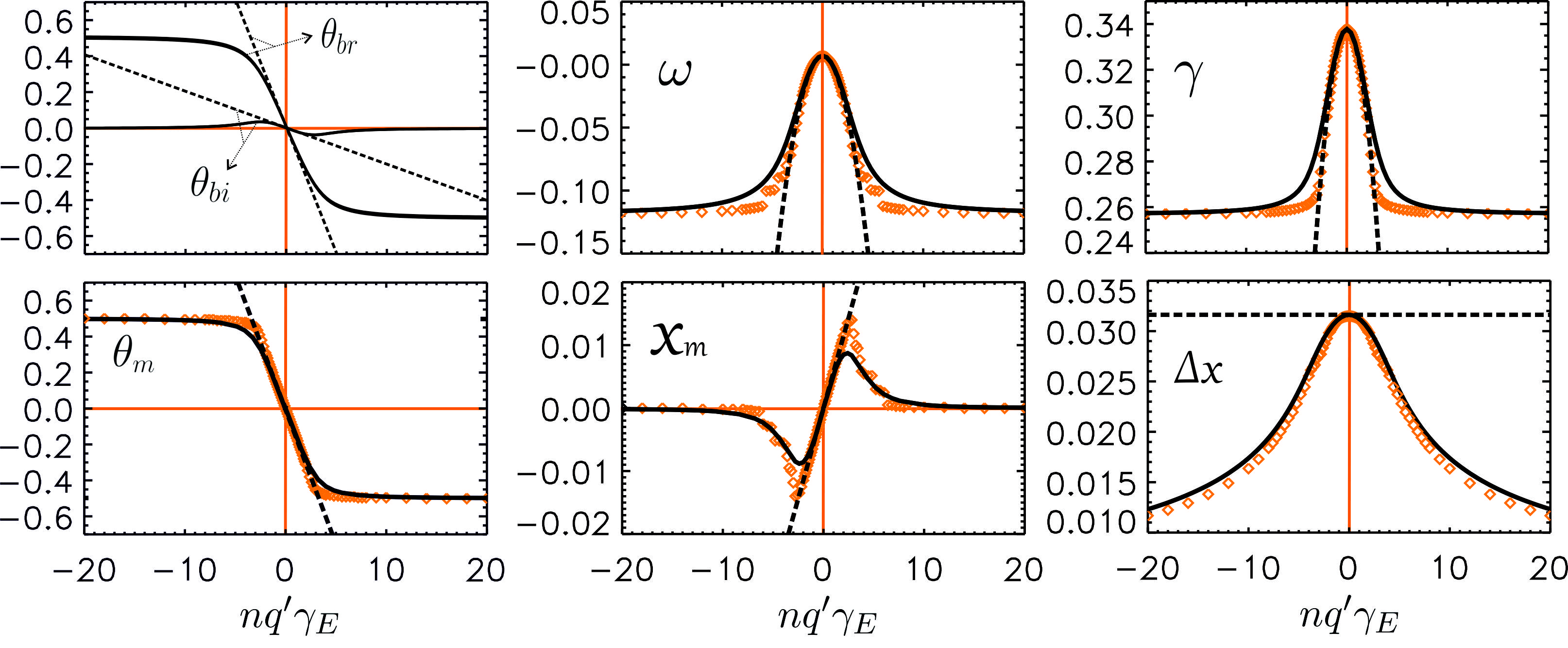}
	\captionof{figure}{\textsf{The global parameters; complex ballooning parameter $\theta_{b} = \theta_{b,r} + i\theta_{b,i}$ (in units of $\pi$), global mode frequency $\Omega = \omega + i\gamma$ (normalised to the electron diamagnetic frequency), the mode physical poloidal position $\theta_{m}$ (in units of $\pi$), the mode physical radial position $x_{m}$ (normalised to the minor radius $r=0.5m$) and finally the mode radial width $\Delta_{x}$ are all plotted as a function of flow shear $\gamma_{E}$ with $n=50$ and $\qprime=10$. Here, we have chosen $\delta=-1$ to obtain a localised mode about $\theta=\theta_{m}$ and sign of $\sigma$ is chosen such that $\sigma=+1$ ($\sigma=-1$) for all values of $\gamma_{E} < 0$ ($\gamma_{E} > 0$). The analytical solutions are represented by solid lines while the square symbols correspond to the numerical solutions. Note that, for comparison, the WABT solutions which are indicated by dashed lines, are also presented.}}\label{fig_1}
\end{figure*}

In this subsection we focus on a mixed mode (MM) that undergoes a small poloidal shift with respect to the outboard (or inboard) mid-plane. Here, we try to reproduce the so-called weakly asymmetric ballooning theory (WABT) of \cite{PeshwazPhD, Xie2016} in the limit $|\lambda_{x}^{2}/\lambda_{1}\lambda_{xx}| \ll 1$. In this limit, treating $|\lambda_{x}^{2}/\lambda_{1}\lambda_{xx}|$ as our small parameter, we apply a so-called regular perturbation theory to solve \Eqn{equ_thetabmodel2} for $\cos{(\theta_{b})}$ and to the first order we obtain  
\begin{eqnarray}\label{equ_MMcosthetab}
\begin{aligned}
\cos{(\theta_{b}^{\pm})} &\approx \pm 1 -\frac{\lambda_{x}^{2}}{2\lambda_{1}\lambda_{xx}}
\end{aligned}
\end{eqnarray}
where the leading order solution of \Eqn{equ_MMcosthetab} is equivalent to \Eqn{equ_IMthetab} corresponding to IMs and anti-IMs. By expanding $\cos{(\theta_{b}^{+})}$ and $\cos{(\theta_{b}^{-})}$ in \Eqn{equ_MMcosthetab} about 0 and $\pm\pi$, respectively, and by retaining up to the second order terms in the expansion, \Eqn{equ_MMcosthetab} leads to the following solutions for $\theta_{b}^{+}$ and $\theta_{b}^{-}$
\begin{eqnarray}\label{equ_plus_MMthetab}
\begin{aligned}
\theta_{b}^{+} \approx \sqrt{\frac{\lambda_{x}^{2}}{\lambda_{1}\lambda_{xx}}}
= \frac{\sigma^{+}\lambda_{x}}{\sqrt{\lambda_{1}\lambda_{xx}}}
\end{aligned}
\end{eqnarray}
and
\begin{eqnarray}\label{equ_minus_MMthetab}
\begin{aligned}
\theta_{b}^{-} \approx \pm\pi + \sqrt{\frac{-\lambda_{x}^{2}}{\lambda_{1}\lambda_{xx}}}
= \pm\pi + \frac{\sigma^{-} i\lambda_{x}}{\sqrt{\lambda_{1}\lambda_{xx}}}
\end{aligned}
\end{eqnarray}
where, sign of $\sigma^{+}=\pm 1$ (or $\sigma^{-}=\pm 1$) controls the mode poloidal shift in the upward or downward direction with respect to the outboard (or inboard) mid-plane.
Now the analytical solutions, using the results given in section \ref{Validation} (from \Eqn{equ_Ymodel} to \Eqn{equ_xmmodel}), are reduced to the following forms\footnote{Here, using the fact that $|\lambda_{x}^{2}/\lambda_{1}\lambda_{xx}| \ll 1$ we have only considered the dominant leading order term in calculating $\delta^{\pm}\sqrt{\frac{\lambda_{1}}{\lambda_{xx}}\cos{(\theta_{b}^{\pm})}}$ from \Eqn{equ_MMcosthetab} and, hence, we have written $\delta^{\pm}\sqrt{\frac{\lambda_{1}}{\lambda_{xx}}\cos{(\theta_{b}^{\pm})}}$ $\approx$ $\delta^{\pm}\sqrt{\pm\lambda_{1}/\lambda_{xx}}$.}:
\begin{eqnarray}\label{equ_MMY}
\begin{aligned}
Y_{M}^{+}(\theta_{0}) \approx \exp{\bigg[-\frac{in\qprime\delta^{+}}{2}\sqrt{\frac{\lambda_{1}}{\lambda_{xx}}} \left(\theta_{0} - \frac{\sigma^{+}\lambda_{x}}{\sqrt{\lambda_{1}\lambda_{xx}}}\right)^{2}\bigg]}
\\ 
Y_{M}^{-}(\theta_{0}) \approx \exp{\bigg[\frac{n\qprime\delta^{-}}{2}\sqrt{\frac{\lambda_{1}}{\lambda_{xx}}} 
\left(\theta_{0} - \frac{\sigma^{-}i\lambda_{x}}{\sqrt{\lambda_{1}\lambda_{xx}}} \mp\pi\right)^{2}\bigg]}
\\ 
\end{aligned}
\end{eqnarray}
\begin{equation}\label{equ_MMom}
\begin{aligned}
\Omega_{M}^{\pm} \approx \lambda_{0} \pm \lambda_{1} - \frac{\lambda_{x}^{2}}{2\lambda_{xx}} + \frac{i\delta^{\pm}\sqrt{\pm\lambda_{1}\lambda_{xx}}}{2n\qprime}
\end{aligned}
\end{equation}
\begin{equation}\label{equ_MMphi}
\begin{aligned}
\tilde{\phi}_{M}^{\pm}(x,\theta) \sim \xi(x, \theta_{m}^{\pm}, \theta) \exp\bigg[-inq_{0}\theta\bigg] 
 \exp\bigg[\frac{in\qprime\delta^{\pm}\sqrt{\pm\lambda_{xx}/\lambda_{1}}}{2} \left(x-x_{m}^{\pm}\right)^{2}\bigg] 
 \exp\bigg[-in\qprime\left(\theta - \theta_{m}^{\pm}\right)x\bigg]
\end{aligned}
\end{equation}
\begin{eqnarray}\label{equ_MMdx}
\begin{aligned}
\Delta_{xM}^{\pm} \approx  \frac{2 \sqrt{2log(2)}}{\sqrt{n|\qprime|}}
\frac{\left|\sqrt{\frac{\pm\lambda_{1}}{\lambda_{xx}}}\right|}
{\sqrt{\left|\left[\sqrt{\frac{\pm\lambda_{1}}{\lambda_{xx}}}\right]_{i}\right|}}
\end{aligned}
\end{eqnarray}
\begin{eqnarray}\label{equ_MMthetam}
\begin{aligned}
\theta_{mM}^{+} &\approx 
\left[\frac{\sigma^{+}\lambda_{x}}{\sqrt{\lambda_{1}\lambda_{xx}}}\right]_{r}
+ \frac{\left[\sqrt{\frac{\lambda_{1}}{\lambda_{xx}}}\right]_{r}  \left[\frac{\sigma^{+}\lambda_{x}}{\sqrt{\lambda_{1}\lambda_{xx}}}\right]_{i}}
		{\left[\sqrt{\lambda_{1}/\lambda_{xx}}\right]_{i}}
\\
\theta_{mM}^{-} &\approx 
\pm\pi - \left[\frac{\sigma^{-}\lambda_{x}}{\sqrt{\lambda_{1}\lambda_{xx}}}\right]_{i}
- \frac{\left[\sqrt{\frac{\lambda_{1}}{\lambda_{xx}}}\right]_{i}  \left[\frac{\sigma^{-}\lambda_{x}}{\sqrt{\lambda_{1}\lambda_{xx}}}\right]_{r}}
{\left[\sqrt{\lambda_{1}/\lambda_{xx}}\right]_{r}}
\end{aligned}
\end{eqnarray}
\begin{eqnarray}\label{equ_MMxm}
\begin{aligned}
x_{mM}^{+} &\approx  \delta^{+} \frac{\left|\sqrt{\frac{\lambda_{1}}{\lambda_{xx}}} \right|^{2} \left[\frac{\sigma^{+}\lambda_{x}}{\sqrt{\lambda_{1}\lambda_{xx}}}\right]_{i}}{\bigg[\sqrt{\frac{\lambda_{1}}{\lambda_{xx}}} \bigg]_{i}} 
\\
x_{mM}^{-} &\approx  \delta^{-} \frac{\left|\sqrt{\frac{\lambda_{1}}{\lambda_{xx}}} \right|^{2} \left[\frac{\sigma^{-}\lambda_{x}}{\sqrt{\lambda_{1}\lambda_{xx}}}\right]_{r}}{\bigg[\sqrt{\frac{\lambda_{1}}{\lambda_{xx}}} \bigg]_{r}}  
\end{aligned}
\end{eqnarray}
where, the subscript \quotes{M} refers to a mixed mode and for $\lambda_{x}=0$ we recapture both IMs and anti-IMs of the previous subsection corresponding to $\theta_{b}^{+}$ and $\theta_{b}^{-}$, respectively. Finally, we note that \Eqn{equ_MMY} to \Eqn{equ_MMxm} are exactly what we would have obtained if we have started with \Eqn{equ_YODE} by neglecting the weak $\theta_{0}$-dependence of both $\lambda_{xx}(\theta_{0})$ and $\lambda_{x}(\theta_{0})$ and expanding about $\theta_{0}=0$ (or $\theta_{0}=\pm\pi$) rather than $\theta_{0}=\theta_{b}$. This is called a weak asymmetric ballooning theory (WABT) which deals only with a MM that sits close to the outboard (or inboard) mid-plane \cite{PeshwazPhD, Xie2016}. Note also that, according to WABT, the radial mode width (from \Eqn{equ_MMdx}) is constant and does not depend on $\lambda_{x}$\footnote{It is worth mentioning that in a parameter regime for which the condition $|\lambda_{x}^{2}/\lambda_{1}\lambda_{xx}| \ll 1$ breaks down, the radial mode width does indeed vary with $\lambda_{x}$. An example of this is a general mode discussed in subsection $\ref{sec_GM}$ (for a more general case see section $\ref{GBT_Application}$, specifically \Fig{fig_1})}.

\subsection{General modes: $\left|\lambda_{x}^{2}/\lambda_{1}\lambda_{xx}\right| \gg 1$}
\label{sec_GM}

We start by applying a so-called singular perturbation theory to solve \Eqn{equ_thetabmodel2} for $|\lambda_{x}^{2}/\lambda_{1}\lambda_{xx}| \gg 1$ and obtain the following two solutions: 
\begin{equation}
\label{equ_plus_GMcosthetab}
\cos(\sigma^{+}\theta_{b}^{+}) \approx + \frac{\lambda_{1}\lambda_{xx}}{\lambda_{x}^{2}} 
\end{equation}
and 
\begin{equation}
\label{equ_minus_GMcosthetab}
\cos(\sigma^{-}\theta_{b}^{-}) \approx -\frac{\lambda_{x}^{2}}{\lambda_{1}\lambda_{xx}}  
\end{equation}
where \Eqn{equ_plus_GMcosthetab} and \Eqn{equ_minus_GMcosthetab} are both symmetric with respect to the sign of $\sigma^{+}=\pm 1$ and $\sigma^{-}=\pm 1$, respectively. We shall first consider the \quotes{$\theta_{b}^{+}$} solution and after substituting $\cos(\sigma^{+}\theta_{b}^{+})$ from \Eqn{equ_plus_GMcosthetab} into \Eqn{equ_alphamodel} and \Eqn{equ_omgmodel}, we obtain
\begin{equation}
\label{equ_plus_GMalpha}
\begin{aligned}
\alpha_{g}^{+} \approx \frac{i\delta^{+}\lambda_{1}}{2\lambda_{x}}
 \end{aligned}
\end{equation}
and 
\begin{equation}
\label{equ_plus_GMomg}
\begin{aligned}
\Omega_{g}^{+} \approx \lambda_{0} + \frac{\lambda_{1}^{2}\lambda_{xx}}{\lambda_{x}^{2}} + \frac{i\delta^{+} \lambda_{1}\lambda_{xx}}{2n\qprime\lambda_{x}},
 \end{aligned}
\end{equation}
respectively. Where, the subscript \quotes{g} refers to a general mode. Note also that if we take the limit $|\lambda_{x}^{2}/\lambda_{1}\lambda_{xx}| \to \pm\infty$, from \Eqn{equ_plus_GMcosthetab} we have $\cos(\theta_{b}^{+}) \to 0$. This leads to two solutions with $\theta_{b}^{+} \to \sigma^{+} \pi/2$, where $\sigma^{+}=\pm 1$, \quotes{plus} and \quotes{minus} signs correspond to a mode that asymptotically approaches the top and bottom of the tokamak plasma, respectively. In this limit, from \Eqn{equ_plus_GMomg} we have\footnote{Note that we can write $\lambda_{0} = \frac{1}{2\pi} \oint{[\lambda(0,\theta_{0})] d\theta_{0}}$, where from \Eqn{equ_lambdamodel} we have $\lambda(0,\theta_{0}) = \lambda_{0} + \lambda_{1}\cos{(\theta_{0})}$  and $\frac{1}{2\pi} \oint[...]d\theta_{0}$ refers to an average over a period in $\theta_{0}$.}
\begin{equation}
\label{equ_plus_GMomg0}
\begin{aligned}
\Omega_{g}^{+} \approx \lambda_{0} 
 \end{aligned}
\end{equation}
This indicates that the mode has a similar eigenvalue as a GM that has been previously studied in the literature, see \cite{Connor1993, Taylor1996,Xie2012} for instance. However, this is necessary but not sufficient to see if they represent the same mode. Therefore, to check this we proceed further with our calculations and investigate the corresponding envelope $Y^{+}(\theta_{0})$. Now using $\alpha_{g}^{+}$ from \Eqn{equ_plus_GMalpha} and after multiplying the right hand side of \Eqn{equ_Ymodel} by a constant factor $\exp{\left[in\qprime\sigma^{+}\delta^{+}\lambda_{1}/\lambda_{x}\right]}$, we obtain the following formula for $Y^{+}(\theta_{0})$
\begin{equation}
\label{equ_plus_GMYth0}
\begin{aligned}
Y_{g}^{+}(\theta_{0}) \approx \exp{\left[in\qprime\frac{\sigma^{+}\delta^{+}\lambda_{1}}{\lambda_{x}} \left(1-\frac{\left(\theta_{0} - \theta_{b}^{+}\right)^{2}}{2}\right)\right]}
 \end{aligned}
\end{equation}
Knowing that the limit $|\lambda_{x}^{2}/\lambda_{1}\lambda_{xx}| \to \pm\infty$ leads to $\theta_{b}^{+} \to \sigma^{+}\pi/2$, the term $\sigma^{+}\left[1-\frac{(\theta_{0} - \theta_{b}^{+})^{2}}{2}\right]$ in the exponent on right hand side of \Eqn{equ_plus_GMYth0}, in turn, can be approximated by $\sigma^{+}\left[1-\frac{\left(\theta_{0} -(\sigma^{+}\pi/2) \right)^{2}}{2}\right]$. This represents a Taylor series expansion of $\sin{(\theta_{0})}$ about $\sigma^{+}\pi/2$ such that \Eqn{equ_plus_GMYth0} is rewritten to get
\begin{equation}
\label{equ_plus_GMYsinth0}
\begin{aligned}
Y_{g}^{+}(\theta_{0}) \approx \exp{\left[ in\qprime\frac{\delta^{+}\lambda_{1}}{\lambda_{x}} \sin{(\theta_{0})}\right]}
 \end{aligned}
\end{equation}
This is similar to the eigenfunction of a GM that one obtains by solving a first order ODE after neglecting the coefficient associated with the second order derivative term in \Eqn{equ_YODE}. Therefore, we have demonstrated that our GBT has captured a general like mode (GM) in the limit $|\lambda_{x}^{2}/\lambda_{1}\lambda_{xx}| \to \pm\infty$. Furthermore, because $\theta_{bi}^{+}=0$, corresponding to $x_{m}=0$ in \Eqn{equ_xmmodel}, we rewrite \Eqn{equ_phixmmodel} to obtain the reconstructed global mode structure $\tilde{\phi}(x,\theta)$ for a GM which reads
	\begin{eqnarray}
\label{equ_plus_GMphi}
\begin{aligned}
\tilde{\phi}_{g}^{+}(x,\theta) \sim \xi(x, \theta_{m}^{+}\approx\sigma^{+} \frac{\pi}{2}, \theta) \exp\bigg[-inq_{0}\theta\bigg] 
 \exp\bigg[\frac{in\qprime\delta^{+}\lambda_{x}}{2\lambda_{1}} x^{2}\bigg] 
\exp\bigg[-in\qprime\left(\theta -\sigma^{+} \frac{\pi}{2}\right)x\bigg]
\end{aligned}
\end{eqnarray}
and the corresponding mode radial width from \Eqn{equ_dxmodel} reduces to
	\begin{eqnarray}
\label{equ_plus_GMdx}
\Delta_{xg}^{+} \approx \frac{2 \sqrt{2log(2)} |\lambda_{1}/\lambda_{x}|}{\sqrt{n|\qprime|}\sqrt{|\left[\lambda_{1}/\lambda_{x}\right]_{i}|}}
\end{eqnarray}
where, from \Eqn{equ_plus_GMdx}, we note that the mode radial width is affected by $\lambda_{x}$, but, according to \Eqn{equ_plus_GMomg0}, its growth rate is unchanged. 

Finally, we investigate the \quotes{$\theta_{b}^{-}$} solution by decomposing \Eqn{equ_minus_GMcosthetab} into its real and imaginary components to obtain
\begin{equation} \label{equ_minus_GMcosthetabri}
\begin{aligned}
		\cos(\sigma^{-}\theta_{br}^{-}) \cosh(\sigma^{-}\theta_{bi}^{-}) & \approx - \left[\frac{\lambda_{x}^{2}}{\lambda_{1}\lambda_{xx}}\right]_{r}
		\\
		\sin(\sigma^{-}\theta_{br}^{-}) \sinh(\sigma^{-}\theta_{bi}^{-}) & \approx 
		\left[\frac{\lambda_{x}^{2}}{\lambda_{1}\lambda_{xx}}\right]_{i}
\end{aligned}
\end{equation}
We shall now square the real and imaginary parts in \Eqn{equ_minus_GMcosthetabri}, replacing $\sinh^{2}{(\sigma^{-}\theta_{bi}^{-})}$ by $\cosh^{2}{(\sigma^{-}\theta_{bi}^{-})} -1$, and add them to obtain $\cosh^{2}{(\sigma^{-}\theta_{bi}^{-})} = \sin^{2}{(\sigma^{-}\theta_{br}^{-})}+\left|\lambda_{x}^{2}/\lambda_{1}\lambda_{xx}\right|^{2}$. Knowing that $0 \le \sin^{2}{(\sigma^{-}\theta_{br}^{-})}\le 1$ and due to the fact that $|\lambda_{x}^{2}/\lambda_{1}\lambda_{xx}| \gg 1$ we conclude that $
\cosh{\sigma^{-}\theta_{bi}^{-}} \gg 1$. This indicates that $|\theta_{bi}^{-}|\gg 1 $. The reconstructed global mode, in turn, undergoes a big radial shift away from the reference rational surface at $x=0$ invalidating the Taylor expansion; therefore, this is not an acceptable solution.

\section{GBT vs. numerical solutions: ITG modes in tokamaks}
\label{GBT_Application}
In this section to benchmark our analytical results against the numerical solutions, we shall consider available data from the literature for a simplified fluid model of ITG modes in large aspect ratio circular tokamaks \cite{DavidPhD, David2014}. To reconstruct the global mode structures with their associated global mode frequencies, local code calculations are usually performed for a range of radial flux surfaces, $x$, and a full $2\pi$ period in $\theta_{0}$, to map out $\lambda(x,\theta_{0})$. Furthermore, in \cite{DavidPhD, David2014}, they have assumed $\lambda_{x}(0, \theta_{0})=\lambda_{x}$ and $\lambda_{xx}(0, \theta_{0})=\lambda_{xx}$ to be constant and, to model the periodic function $\lambda(0, \theta_{0})$, only two Fourier modes are retained. This is the same model that we have considered in Section \ref{Validation}. The coefficients $\lambda_{xx}$, $\lambda_{0}$ and $\lambda_{1}$ have the following numerical values
\begin{equation} \label{equ_benchmark_0}
\begin{split}
\lambda_{xx} &= 25.2200 - i 31.8000 
\\
\lambda_{0}  &= -0.1183 + i 0.2571 
\\
\lambda_{1} &=  0.1257 + i 0.0831 
\end{split}
\end{equation}
In addition, we incorporate flow shear via Doppler shift in $\lambda_{x}$ and write
\begin{equation}
\label{equ_benchmark_1}
\begin{aligned}
\lambda_{xr} &= n\qprime\gamma_{E}
\\
\lambda_{xi} &= 0
\end{aligned}
\end{equation}
where the flow shearing rate
\begin{equation}
\label{equ_benchmark_2}
\begin{aligned}
\gamma_{E} = \frac{d\Omega_{\phi}}{dq} = \frac{1}{\qprime}\frac{d\Omega_{\phi}}{dx} 
\end{aligned}
\end{equation}
is constant and $\Omega_{\phi}$ is the toroidal rotational frequency of the magnetic flux surfaces with respect to the reference surface at $x=0$. Note that, for this model, we have $[\lambda_{1}]_{i} = 0.0831 > 0$ implying, according to \Eqn{equ_IMDelom}, that an IM on the outboard mid-plane is more unstable than an anti-IM on the inboard mid-plane. Moreover, solving \Eqn{equ_thetabmodel2} for $\theta_{b}$ leads to the following two solutions:
\begin{equation}
\label{equ_thetabmodelflow}
\theta_{b}^{\pm} = \sigma^{\pm} \cos^{-1}{\left[ -\frac{n^{2}\qprime^{2}\gamma_{E}^{2}}{2\lambda_{1}\lambda_{xx}} 
\pm \sqrt{\left(\frac{n^{2}\qprime^{2}\gamma_{E}^{2}}{2\lambda_{1}\lambda_{xx}}\right)^{2} + 1} \right]}
\end{equation}
where in the good curvature region, corresponding to the \quotes{minus} sign solution, as $\gamma_{E}$ increases the mode undergoes a big radial shift with respect to $x=0$. However, the mode in the bad curvature region, corresponding to \quotes{plus} sign solution stays localised close to $x=0$ for all values of $\gamma_{E}$. Hence, in \Fig{fig_1} we only explore the \quotes{plus} sign solutions.

Using the model coefficients from \Eqn{equ_benchmark_0} and \Eqn{equ_benchmark_1}, \Fig{fig_1} shows how the complex ballooning parameter $\theta_{b}$, global mode frequency $\Omega$, mode physical poloidal position $\theta_{m}$, mode physical radial position $x_{m}$ and the mode radial width $\Delta_{x}$ vary as functions of $\gamma_{E}$. Analytical solutions corresponding to GBT are derived in \Eqn{equ_omgmodel}, \Eqn{equ_dxmodel}, \Eqn{equ_thetammodel}, \Eqn{equ_xmmodel} and \Eqn{equ_thetabmodel2} (Note that we have solved \Eqn{equ_thetabmodel2} for $\theta_{b}$ and the result is presented in \Eqn{equ_thetabmodelflow}) while the WABT solutions are obtained from \Eqn{equ_plus_MMthetab}, \Eqn{equ_MMom}, \Eqn{equ_MMdx}, \Eqn{equ_MMthetam} and \Eqn{equ_MMxm}, respectively\footnote{Note that, to obtain a localised mode about $\theta=\theta_{m}$ we have chosen $\delta=-1$. In addition, sign of $\sigma$ is chosen such that $\sigma=+1$ and $\sigma=-1$ corresponding to $\gamma_{E} < 0$ and $\gamma_{E} > 0$, respectively.}. As we can see from \Fig{fig_1}, for small values of flow shear corresponding to $|n\qprime\gamma_{E}| < 3.0$, both GBT solutions (solid lines) together with the WABT solutions (dashed lines), are in excellent agreement with the data obtained from the full numerical solutions of \Eqn{equ_YODE} (squared symbols). However, as flow shear increases beyond this region, the WABT solutions break down and cannot reproduce the numerical solutions. This is expected, because WABT only works for small poloidal shifts with respect to the outboard mid-plane. On the contrary, the GBT solutions correctly reproduce the numerical solutions over the full range of flow shear. Finally, in what follows, we shall only use the GBT solutions, considering an IM, MM and GM separately, and examine in detail the corresponding reconstructed global mode structures with their stability properties.

\begin{figure*}[!t]
	\centering
	\includegraphics[width=5.0in]{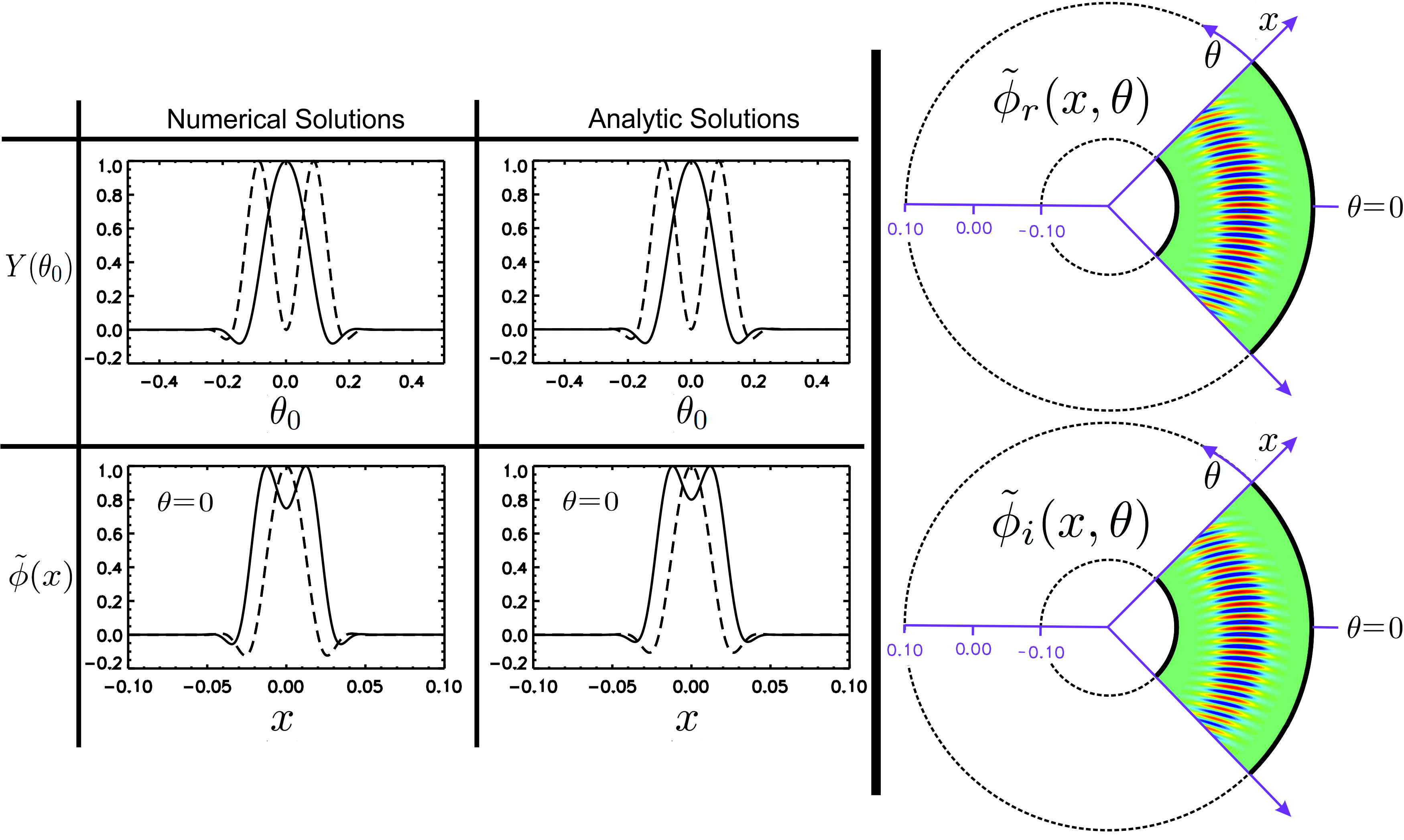}
	\caption{\textsf{The envelope $Y(\theta_{0})$ and the associated reconstructed global mode structure $\tilde{\phi}(x,\theta)$ for $n\qprime\gamma_{E}=0$ (corresponding to an IM). [left] Presents both analytic and numerical solutions for $Y(\theta_{0})$ and $\tilde{\phi}(x,\theta=\theta_{m}=0)$ as a function of $\theta_{0}$ and $x$, respectively. Here, $\theta_{0}$ is measured in units of $\pi$ while $x$ is normalised to the minor radius $r=0.5m$. Note that the solid and dashed lines correspond to the real and imaginary components, respectively. [right] The color contour plot of $\tilde{\phi}(x,\theta) = \tilde{\phi}_{r}(x,\theta) + i \tilde{\phi}_{i}(x,\theta)$ in the poloidal cross-section. Note that, the mode has radial symmetry.}}
	\label{fig_2}
\end{figure*}
\begin{figure*}[!t]
	\centering
	\includegraphics[width=5.0in]{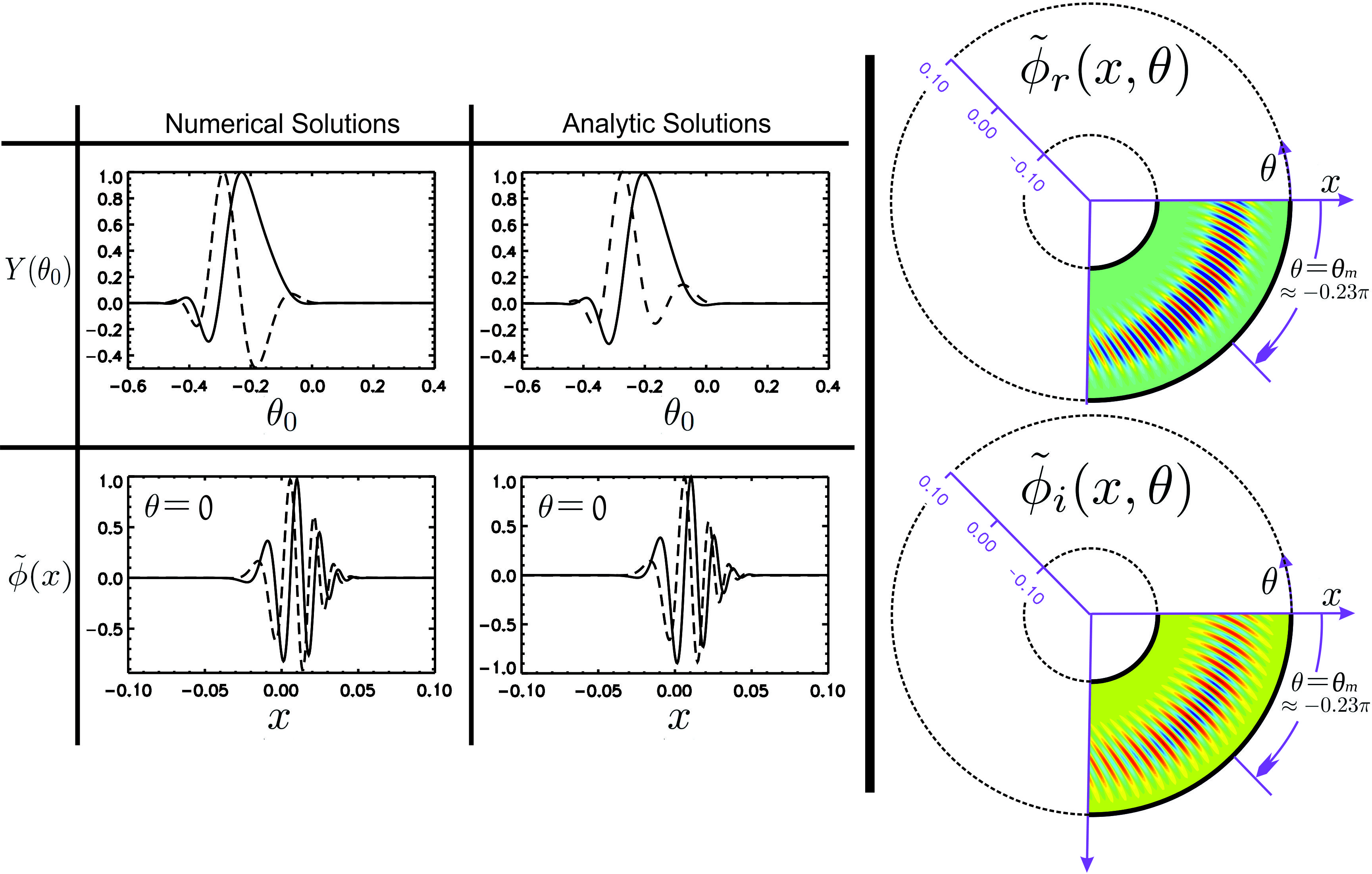}
	\caption{\textsf{Equivalent set of plots to those presented in \Fig{fig_2}, but with $n\qprime\gamma_{E}=1.5$ which leads to a global mode that peaks poloidally at $\theta=\theta_{m} \approx -0.23\pi$, rather than $\theta=\theta_{m}=0$, and, in turn, has radial asymmetry.}}
	\label{fig_3}
\end{figure*}
\begin{figure*}[!t]
	\centering
	\includegraphics[width=5.0in]{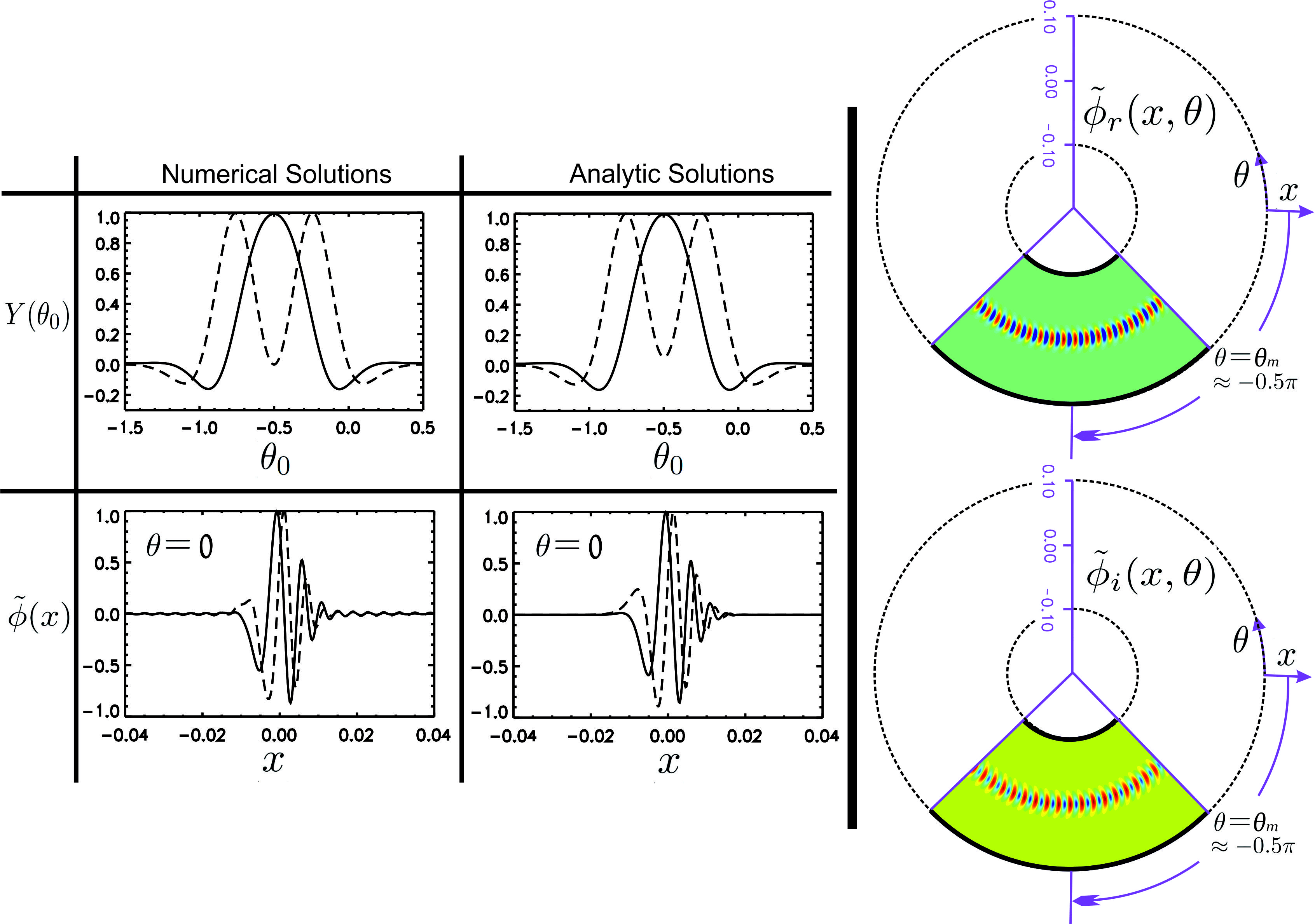}
	\caption{\textsf{Equivalent set of plots to those presented in \Fig{fig_2}, but with $n\qprime\gamma_{E}=20$ which leads to a general mode that sits at the bottom of the tokamak plasma, i.e $\theta=\theta_{m}\approx -0.5\pi$, and, in turn, has radial asymmetry.}}
	\label{fig_4}
\end{figure*}

For $\gamma_{E} = 0$, using our GBT analytical solutions from section \ref{sec_IM} we capture an IM for which $\theta_{b} = \theta_{m} = x_{m} = 0$. Furthermore, from \Eqn{equ_IMY} and \Eqn{equ_IMom} we obtain the analytical solutions for both the envelope $Y(\theta_{0})$ and the associated global mode frequency $\Omega = 0.0070+i 0.3378$. Moreover, the reconstructed global mode structure $\tilde{\phi}(x,\theta)$ with its radial width $\Delta_{x} = 0.0316$ are calculated from \Eqn{equ_IMphi} and \Eqn{equ_IMdx}, respectively. The corresponding numerical solutions are $\Omega = 0.0070+i 0.3378$ and $\Delta_{x} = 0.0316$. Here, we have solved \Eqn{equ_YODE} numerically for both $Y(\theta_{0})$ and $\Omega$ and, to evaluate $\Delta_{x}$, we have substituted the numerical data for $Y(\theta_{0})$ into the Fourier-ballooning representation in \Eqn{equ_FB}. Numerical integration of \Eqn{equ_FB}, in turn, provides $\tilde{\phi}(x,\theta)$ and, finally, we defined $\Delta_{x}$ as the full width at half maximum of a Gaussian fit to the magnitude of $\tilde{\phi}$ at $\theta = \theta_{m}=0$. In \Fig{fig_2}, on the left hand side, the envelope $Y(\theta_{0})$ and the associated reconstructed global mode structure $\tilde{\phi}(x,\theta=\theta_{m}=0)$, both the real (solid lines) and imaginary (dashed lines) components, are benchmarked against their full numerical solutions. On the right hand side of the same figure, the color contour plots of $\tilde{\phi}(x,\theta) = \tilde{\phi}_{r}(x,\theta) + i \tilde{\phi}_{i}(x,\theta)$ in $x$ and $\theta$ space are presented. It is quite clear that the envelope $Y(\theta_{0})$ is symmetric about $\theta_{0}=0$ leading to a reconstructed global mode that peaks on the outboard mid-plane, radially centered on $x=x_{m}=0$, and has a radial symmetry. This is exactly the case where $\lambda(x,\theta_{0})$ has a stationary point at $x=0$, leading to an IM that has been previously studied in \cite{Connor1978, Taylor1996, David2014} for example (see also our section \ref{sec_IM} for more details). Finally, excellent agreement between our analytic solutions and the numerical ones are found.

Furthermore, as we switch on the flow shear the mode shifts away from the outboard mid-plane and, hence, its radial symmetry is broken. To illustrate this symmetry breaking that is associated with this class of eigenmode solutions, namely mixed modes, we consider an additional effect from $n\qprime\gamma_{E}=1.5$. This corresponds to a local complex mode frequency $\lambda(x,\theta)$ that does not exhibit a stationary point on the real $x$ axis, which in turn, according to \Eqn{equ_thetammodel} leads to $\theta_{m} = -0.224\pi$ and generates a global mode that poloidaly shifts downward with respect to the outboard mid-plane. \Fig{fig_3} shows an equivalent set of plots to those presented in \Fig{fig_2}, but here instead of $n\qprime\gamma_{E}=0$ we have $n\qprime\gamma_{E}=1.5$. The calculated complex ballooning parameter $\theta_{b} = -0.1907-i0.0304$, mode radial position $x_{m}=0.0086$, the mode radial width $\Delta_{x}=0.0307$ and finally global mode frequency $\Omega = -0.0102 + i0.3160$ are obtained from \Eqn{equ_thetabmodel2} (or equivalently \Eqn{equ_thetabmodelflow}), \Eqn{equ_xmmodel}, \Eqn{equ_dxmodel} and \Eqn{equ_omgmodel}, respectively. The corresponding numerical solutions are $\theta_{m} = -0.230\pi$, $x_{m} = 0.0087$, $\Delta_{x} = 0.0307$ and $\Omega = -0.0104 + i0.3162$, respectively. Moreover, the envelope $Y(\theta_{0})$ is not symmetric about $\theta_{0} = \theta_{m} = -0.23\pi$ (corresponds to $\theta_{bi} = -0.0304 \ne 0$) and this is the source of the above observed radial shift ($x_{m} \ne 0$). In addition, because $\theta_{m} \ne 0$, the reconstructed global mode structure is not radially symmetric about $x=x_{m}$. We also point out that in this region, corresponding to the intermediate values of flow shear, a slight discrepancy between our GBT results and numerical solutions are observed. This might be attributed to the fact that we have only retained the first two terms when we Taylor expanded \Eqn{equ_YODE} about $\theta_{0}=\theta_{b}$.

Finally, as we can see for large values of flow shear, $n\qprime\gamma_{E} \gg 1$, the global parameters in \Fig{fig_1} approach constant values. For this limit, considering $n\qprime\gamma_{E}=20$, \Fig{fig_4} shows the envelope $Y(\theta_{0})$ together with the corresponding reconstructed global mode structure $\tilde{\phi}(x,\theta)$. Here, to obtain the numerical solutions we have solved \Eqn{equ_FB} and \Eqn{equ_YODE} numerically, while the corresponding analytic solutions for $Y(\theta_{0})$ and $\tilde{\phi}(x,\theta)$ are given by \Eqn{equ_Ymodel} and \Eqn{equ_phixmmodel} (or equivalently\Eqn{equ_plus_GMYsinth0} and \Eqn{equ_plus_GMphi}), respectively. Note that, due to the fact that $Y(\theta_{0})$ is symmetric about $\theta_{0}=-\pi/2$, corresponding to $\theta_{bi} = 0$, $\tilde{\phi}(x,\theta)$, in turn, is centred on $x=0$. We can see that the radial slice from the constructed eigenmode structure is not symmetric on the outboard mid-plane at $\theta=0$. Moreover, the analytic solutions for $\Omega = -0.1183 + i0.2571$ and $\Delta_x = 0.0122$ are obtained from \Eqn{equ_omgmodel} and \Eqn{equ_dxmodel} (or equivalently \Eqn{equ_plus_GMomg0} and \Eqn{equ_plus_GMdx}), respectively, and their corresponding numerical solutions are $\Omega = -0.1179 + i 0.2571$ and $\Delta_x = 0.0117$. Finally, it is worth mentioning that, good agreement is again found between analytical and numerical solutions.

\section{Summary and conclusions}
\label{conclusions}
In the context of higher order ballooning theory, we have developed a new analytical approach, namely generalised ballooning theory (GBT), to account for a global mode that balloons at arbitrary poloidal position with respect to the outboard (or inboard) mid-plane of tokamak plasmas. The so-called weak asymmetric ballooning theory (WABT), which only provides accurate results for a mode that sits close to the outboard mid-plane, is recaptured as a special limit from our GBT theory. Furthermore, to benchmark GBT against the numerical solutions, we have used available data from a simplified fluid model for the toroidal linear electrostatic ITG modes in circular tokamaks. It is found that the parameter regimes with moderate and large poloidal shifts, corresponding to $|\gamma_{E}| > 3$ in \Fig{fig_1}, cannot be covered in the context of WABT treatment, but GBT works well in this region. Specifically, the inter-mode transition between IM and GM can only be captured in the context of our GBT theory.  

Moreover, it is shown that the imaginary component of the ballooning parameter $\theta_{b}$ leads to a mode that radially shifts away from its associated rational surface at $x=0$. In addition, our theory has revealed that the IMs are the only modes that preserve radial symmetry. The radial asymmetry that is associated with the non isolated modes, caused by profile shearing, can be used, in the context of GBT, as an important tool to investigate the mechanism that generates the so-called Reynolds stress. 

Furthermore, GBT is a quite general theory which can be employed and combined with a state-of-art local gyrokinetic code, GS2, GKW, GENE and GYRO for instance, to improve our understanding of the linear microinstabilities for realistic experimental regimes. Unlike previous ballooning treatments, our theory can be applied to experimentally relevant equilibria for which the shape of magnetic flux surfaces are not, in general, circular and one usually needs to consider the effect of radial profile variations. Finally, it is worth mentioning that our GBT theory, as with previous ballooning treatments, has the following limitations: \textbf{(a)} \Eqn{equ_YODE} is derived under the assumption that equilibrium quantities vary slowly across rational surfaces, \textbf{(b)} the approach only works in linear regimes. It will be interesting to consider whether this approach can be extended to the non-linear regimes, but this is quite challenging and it is not clear yet how to achieve this. Nonetheless, the linear instabilities with their mode structures remain interesting; for example, quasi-linear theory, frequently used to model the transport that is associated with microturbulence, assumes that the saturated non-linear mode structure continues to resemble the linear mode. 

\section*{Acknowledgment}
P. A. Abdoul appreciate the financial support by the Ministry of Higher Education in Kurdistan region of Iraq. This work has been (part-) funded by the RCUK Energy Programme [grant number EP/P012450/1].




\end{document}